\documentclass[prd, aps, superscriptaddress, preprintnumbers, twocolumn, floatfix, nofootinbib]{revtex4-2}
\pdfoutput=1

\usepackage{amsfonts}
\usepackage{amsmath}
\usepackage{amssymb}
\usepackage{bm}
\usepackage{dcolumn}
\usepackage{graphicx}
\usepackage[latin1]{inputenc}
\usepackage{latexsym}
\usepackage{rotating}
\usepackage{hyperref}
\usepackage{graphicx}
\usepackage{xcolor}

\newcommand\be{\begin{equation}}
\newcommand\bea{\begin{eqnarray}}
\newcommand\ee{\end{equation}}
\newcommand\eea{\end{eqnarray}}

\def\doi{http://doi.org}

\begin{document}

\title{A Bouncing Cosmology from VECROs}

\author{Robert Brandenberger}
\email{rhb@physics.mcgill.ca}
\affiliation{Department of Physics, McGill University, Montr\'{e}al, QC, H3A 2T8, Canada}


\author{Gabrielle A. Mitchell}
\email{gabrielle.mitchell@mail.mcgill.ca}
\affiliation{Department of Physics, McGill University, Montr\'{e}al, QC, H3A 2T8, Canada}

\date{\today}

\begin{abstract} 
 
We argue that, in the same way that in a black hole space-time VECROs will form in order to cancel the gravitational effects of a collapsing mass shell and prevent the formation of a singularity,  in a contracting universe a gas of VECROs will form to hold up the contraction, prevent a Big Crunch singularity, and lead to a nonsingular cosmological bounce.

\end{abstract}

\pacs{98.80.Cq}
\maketitle

\section{Introduction}

The {\it fuzzball} program (see \cite{Mathur0} for reviews) is an approach to understand the microscopic origin of the entropy of black holes, and also to provide a solution of the {\it Black Hole Information Loss} problem.  Fuzzballs are the microstates inside of the black hole, and the number of possible microstates is argued to lead to the famous Beckenstein formula for the entropy of a black hole \cite{BHentropy}.  The fuzzball states have support everywhere inside the black hole horizon. In this approach, the individual black hole microstates do not have a horizon.  The horizon arises at the level of the effective theory after coarse graining.  Hence, information can escape from the black hole, and there is no black hole information loss problem. Note that fuzzballs carry positive energy.

More recently,  the concept of VECROs (Virtual Extended Compression-Resistant Objects) has been introduced \cite{Mathur1} to understand what happens when a shell of matter collapses to form a black hole.  The idea is that when a shell of regular matter collapses and falls into its Schwarzschild radius, the vacuum responds by generating  a set of virtual extended objects stretching out beyond the location of the mass shell. VECRO formation is closely related to fuzzball formation, in a way which we will come back to later on. VECROs are also argued to cap off the geometry at the center of the mass concentration, thus eliminating the singularity of the metric.
 
 In this short note we speculate that VECROs will also form in a contracting cosmology and will resist the contraction to a singularity.  Rather, they will mediate a nonsingular cosmological bounce.

Key to reaching our conclusion is the realization that, in order for VECROs to eliminate the black hole singularity,  the effective energy density with which they couple to gravity needs to be negative.  We then argue that a gas of VECROs in a contracting cosmology will have the equation of state $p = \rho$ of a stiff fluid \footnote{$\rho$ and $p$ denote energy density and pressure, respectively.}.  Combining these two arguments, the Friedmann equations then immediately yield a nonsingular bounce. 

Our work is motivated in part by proposals \cite{BF, Ven, Jerome} that the early universe could be described by a gas of black holes \cite{BF} or ``string holes'' \cite{Ven, Jerome} which has a stiff equation of state.

\section{Key Arguments}

We will make use of two key arguments. The first is that, in order to resolve the black hole singularity in the context of the effective field theory of Einstein gravity, the VECRO energy density must be negative.  To show this, we start with the metric of a vacuum black hole without charge and without rotation
\be
ds^2 \, = \, - (1 - \frac{2GM}{r}) dt^2 + (1 - \frac{2GM}{r})^{-1} dr^2
\ee
(we drop the angular variables), where $t$ is the time of an observer far from the center of the black hole, and $r$ is the radial distance from the hole,  chosen such that the area of the sphere surrounding the hole at radius $r$ is $4 \pi r^2$. In the presence of matter, the metric of a static, spherically symmetric space-time can be written in the form
\be
ds^2 \, = \, - A(r) dt^2 + B(r) dr^2 \, ,
\ee
with coefficient functions $A(r)$ and $B(r)$ which are determined by the distribution of matter. If we use the parametrization
\be
A(r) \, \equiv \, e^{2 h(r)} f(r) 
\ee
and
\be
B(r) \, \equiv \, f(r)^{-1} \, ,
\ee
then the Einstein equations lead to the following equations (see e.g. \cite{Blau}) which determine the coefficient functions in terms of the components of the energy-momentum tensor $T_{\mu \nu}$:
\be \label{E1}
h^{\prime}(r) \,  = \, 4 \pi G r f(r)^{-1} \bigl( {T^r}_r - {T^t}_t \bigr)
\ee
and
\be \label{E2}
\bigl[ r ( f(r) - 1 ) \bigr]^{\prime} \, = \, 8 \pi G r^2 ( - {T^t}_t ) \, ,
\ee
where a prime denotes a derivative with respect to $r$.  Redefining
\be
r ( f(r) - 1 ) \, \equiv \, - 2 m(r) 
\ee
such that
\be
f(r) \, = \, 1 - \frac{2 m(r)}{r} \, ,
\ee
Eq. (\ref{E2}) becomes
\be
m(r)^{\prime} \, = \, 4 \pi G r^2 (- {T^t}_t ) \, = \, 4 \pi G r^2 \rho(r) 
\ee
and hence
\be
m(r) \, = \, 4 \pi G \int_0^r dr' r'^2 \rho(r') \, .
\ee

Let us now consider a shell of mass $m_0$ collapsing to form a black hole, and assume that the mass shell is not held up by some pressure. . Then in order to avoid the singularity at $r = 0$, it is necessary that a compensating energy component with negative energy density $\rho_v(r) < 0$ builds up just outside of the collapsing shell. This is the ``VECRO'' component.    
 \be
 m(r) \, = \, m_0 + 4 \pi G \int_0^r dr^{\prime} (r^{\prime})^2 \rho_v(r^{\prime}) \, .
 \ee

 If $r_m$ is the radius of the collapsing shell, then the VECRO component has to be localized just outside of $r_m$, and its contribution to the mass must cancel $m_0$ at the radius $r_m$. This implies that, as $r_m$ tends to zero, the VECRO energy density diverges. For $r > r_s$ (where $r_s$ is the Schwarzschild radius associated with the mass $m_0$) we require $m(r) = m_0$. This implies that there is a positive energy density component $\rho_f(r)$ which at values $r \rightarrow r_s$ cancels the negative energy contribution of the VECROs. This component is the fuzzball component \cite{Mathur0}.  Note that $\rho_f(r)$ has support between $r_m$ and $r_s$, and does not diverge as $r_m$ tends to zero \footnote{This discussion illustrates the relationship between VECRO and fuzzball formation.}.
 
The second key argument in our analysis is the claim that the equation of state of a gas of VECROs in a contracting universe is $w = 1$, i.e.
 \be \label{EoS}
 p_v \, = \, \rho_v \, .
 \ee
This is the same equation of state as that of the black hole gas discussed in \cite{BF} and the gas of string holes analyzed in \cite{Ven}. To demonstrate this,we can follow the discussion in \cite{Jerome}. The crucial input in the analysis of \cite{Jerome} is that the entropy $S$ of a gas of VECROs is given in terms of the mass of a single VECRO in the same way as for a black hole, namely holographically
\be
S \, = \, N M_v^2 G \, ,
\ee
where $M_v$ is the mass of a single VECRO and $N$ is the number of these objects in some fixed comoving volume $V$.  The energy $E$ of a gas of VECROs is
\be
E \, = \, N M_v \, .
\ee
Using the thermodynamic identities
\be
P \, = \, T \bigl( \partial S / \partial V \bigr)_E 
\ee
(where $P$ is the total pressure and $T$ is the temperature) and
\be
T^{-1} \, = \bigl( \partial S / \partial E \bigr)_V
\ee
we then immediately obtain (\ref{EoS}).

 \section{Vecros and a Nonsingular Bounce}

In analogy to how we have argued in the previous section that the negativity of the VECRO energy density can cap off the singularity at the center of a collapsing mass shell, we now argue that the negative energy density of VECROs, combined with their stiff equation of state, will naturally lead to a nonsingular cosmology.

We consider a homogeneous and isotropic contracting universe filled with cold matter (equation of state $p = 0$), radiation (equation of state $p = 1/3 \rho$) and vecros (equation of state $p = \rho$).  To be specific, we will consider a spatially flat universe. From the respective continuity equations, it immediately follows that the energy density in cold matter scales as $\rho_m(t) \sim a(t)^{-3}$, where $a(t)$ is the cosmological scale factor,  that in radiation as $\rho_r(t) \sim a^{-4}$, and that in VECROs as $\rho_v \sim a^{-6}$. As we approach $a(t) = 0$, radiation will dominate over cold matter, and thus in the following we will neglect the cold matter component. 

The Friedmann equation reads
\be \label{FRW1}
H^2 \, = \, \frac{8 \pi G}{3} \rho \, ,
\ee
where $H \equiv {\dot{a}}/a$ is the Hubble expansion rate,  and $G$ is Newton's gravitational constant, and
\be \label{FRW2}
\frac{\ddot{a}}{a} \, = \, - \frac{4 \pi G}{3} \bigl( \rho + 3 p \bigr) \, .
\ee

We assume that early in the contracting phase the energy density in VECROs 
\be
\rho_v(t) \, = \, - A_v \bigl( \frac{a(t)}{a(t_i)} \bigr)^{-6}\, ,
\ee
where $t_i$ can be taken to be some initial time, and $A_v$ is a positive amplitude, is smaller than the energy density in radiation
\be
\rho_r(t) \, = \, A_r \bigl( \frac{a(t)}{a(t_i)} \bigr)^{-4}\, ,
\ee
where $A_r$ is the initial radiation density amplitude.  As the Universe contracts and $a(t)$ decreases,  the absolute value of the energy density in VECROs will catch up to the energy density in radiation, and they will become equal at some time $t_b$ and we will have (from (\ref{FRW1})) 
\be
H(t_b) \,  = \, 0 \, .
\ee
Since the equation of state $p / \rho$ for VECROs is larger than for radiation, then at the time $t_b$ the abolute value of the VECRO pressure will be larger than the radiation pressure, and by (\ref{FRW2}) it then follows that
\be
{\dot{H}}(t_b) \, > \, 0 \, .
\ee
Thus, we see that the addition of the VECRO component will lead to a smooth cosmological bounce.

\section{Discussion}
 
 In this note we have argued that a gas of VECROs in a contracting univere can yield a nonsingular cosmological bounce. VECROs can be viewed as a quantum response of the vacuum which works to counteract a developing singularity, in the same way that VECROs form in the case of spherical infall to prevent the formation of a Schwarzschild singularity.
 
One key ingredient in our argument is the assumption that the energy density of VECROs is negative when considering its coupling to classical gravity.  We have argued for this based on the role which VECROs play in removing the singularity at the center of a black hole.  We also propose an analogy with the back-reaction of infrared modes of cosmological perturbations \footnote{See \cite{Abramo} for original work, \cite{BRreview} for a review, and \cite{entropy} for studies demonstrating the need for the presence of entropy fluctuations in order to obtain a nonvanishing effect.}. A matter fluctuation will create a gravitational potential well, and on super-Hubble scales the negativity of the gravitational energy overwhelms the positivity of the matter energy, thus leading to a net negative energy density which leads to a reduction of the locally observable expansion rate.  In analogy, space inside a contracting matter shell will respond gravitationally to produce a negative effective energy density which tends to reduce the locally measured curvature inside the black hole.

A uniform contracting cosmology will create a packed gas of VECROS. The holographic entropy scaling of VECROs then leads to the conclusion that the equation of state is that of a stiff fluid $p = \rho$. This is the second key ingredient of our argument. 

These two arguments then immediately imply that a homogeneous collapsing cosmology with undergo a nonsingular bounce because of the formation of a gas of VECROs which resist the contraction.
 
Bouncing cosmologies form a class of interesting resolutions of the Big Bang singularity. Bouncing cosmologies (see e.g. \cite{BounceRev} for a review) can also provide an alternative to cosmological inflation as a solution to the horizon problem of Standard Big Bang Cosmology \footnote{The {\it emergent scenario} is another alternative to cosmological inflation which may naturally arise in string theory, as discussed in the context of {\it String Gas Cosmology} \cite{BV} or, more recently, matrix model cosmology \cite{us} (see also \cite{Vafa}). In this case,  it is thermal fluctuations in the emergent phase which lead to the curvature fluctuations and gravitational waves which are observed today \cite{Nayeri}.}. In the same way that in certain accelerating expanding cosmologies quantum vacuum perturbations can develop into an approximately scale-invariant spectrum of curvature fluctuations \cite{ChibMukh} and gravitational waves \cite{Starob}, in certain classes of decelerating contracting cosmologies quantum vacuum perturbations can also develop into approximately scale-invariant perturbations after the bounce.  One example is the {\it matter bounce} scenario \cite{Fabio} (defined as a model in which the equation of state is that of cold matter during contraction when scales which are currently observed exit the Hubble horizon). This model, however, is unstable towards the development of anisotropies \cite{Peter}. Another possibility is to have a phase of Ekpyrotic contraction (given by an equation of state $w \gg 1$) \cite{Ekp} in which case initial vacuum perturbations evolve into a scale-invariant spectrum of metric fluctuations which can also lead to scale-invariant curvature perturbations after the bounce \cite{Durrer} (see e.g. \cite{Ziwei} for a concrete realization). A key challenge for bouncing cosmologies is how to obtain the cosmological bounce. Here, we are proposing a mechanism which yields such a bounce. In the way we have presented the argument, it works as long as there is no other component of matter with an equation of state $w > 1$. Thus,  the VECRO mechanism will automatically resolve the singularity in the matter bounce scenario. In the case of the Ekpyrotic scenario,  we need to require that the phase of Ekpyrotic contraction ends at sufficiently high energy densitites.  This appears to be a rather mild requirement since before the string energy density is reached, the negative exponential potential of the scalar field which yields Ekpyrotic contraction is expected to flatten out and hence the equation of state of the the scalar field will revert to $w < 1$.

Note that if the energy scale at the bounce point is lower than the Planck scale, then the amplitude of the cosmological fluctuations will remain in the linear regime (if, as is commonly assumed in bouncing cosmoloiges, the fluctuations originate as quantum vacuum perturbations on sub-Hubble scales in the far past). 

The VECRO mechanism which we are proposing in this note addresses a key challenge for bouncing cosmologies. This is all the more important since recent arguments indicate that cosmologies with a phase of sufficient length of accelerated expansion to realize the inflationary scenario are hard to realize in string theory (they are in tension with the ``swampland'' criteria \cite{swamp}) and also face conceptual problems from unitarity considerations (the ``Trans-Planckian Censorship Conjecture'' (TCC) \cite{TCC1, TCC2} (see also \cite{RHBrevs} for reviews)), while bouncing and emergent scenarios are consistent with the TCC as long as the energy density at the bounce or in the emergent phase is smaller than the Planck scale.

A key open issue is a clear derivation of VECROs in cosmology from superstring theory. At a superficial level,  our proposed gas of VECROs is in tension with energy conditions for an effective four space-time dimensional theory derived from string theory \cite{Bernardo}, where it is shown that the Null Energy Condition (NEC) must be obeyed.  While our total energy density obeys the NEC, the gas of VECROs does not since it has negative energy density.  The assumptions made in \cite{Bernardo} on the continuity properties of the scale factor are obeyed in our model, our VECROs are quantum objects, and hence the analysis of \cite{Bernardo} does not directly apply.  Nevertheless, this issue requires further study.

\section*{Acknowledgments}

We are grateful to Heliudson Bernardo, Keshav Dasgupta, Jerome Quintin and (in particular) Samir Mathur for discussions. The research at McGill is supported in part by funds from NSERC and from the Canada Research Chair program.


\begin{thebibliography}{99}

\bibitem{Mathur0}
S.~D.~Mathur,
``Fuzzballs and the information paradox: A Summary and conjectures,''
[arXiv:0810.4525 [hep-th]];\\
I.~Bena, E.~J.~Martinec, S.~D.~Mathur and N.~P.~Warner,
``Fuzzballs and Microstate Geometries: Black-Hole Structure in String Theory,''
[arXiv:2204.13113 [hep-th]].

\bibitem{BHentropy}
J.~D.~Bekenstein,
``Black holes and entropy,''
Phys. Rev. D \textbf{7}, 2333-2346 (1973)
doi:10.1103/PhysRevD.7.2333

\bibitem{Mathur1}
S.~D.~Mathur,
``The VECRO hypothesis,''
doi:10.1142/S0218271820300098
[arXiv:2001.11057 [hep-th]].

\bibitem{BF}
T.~Banks and W.~Fischler,
``Holographic cosmology 3.0,''
Phys. Scripta T \textbf{117}, 56-63 (2005)
doi:10.1238/Physica.Topical.117a00056
[arXiv:hep-th/0310288 [hep-th]];\\
T.~Banks and W.~Fischler,
``Holographic cosmology,''
[arXiv:hep-th/0405200 [hep-th]];\\
T.~Banks and W.~Fischler,
``The holographic approach to cosmology,''
[arXiv:hep-th/0412097 [hep-th]].

\bibitem{Ven}
G.~Veneziano,
``A Model for the big bounce,''
JCAP \textbf{03}, 004 (2004)
doi:10.1088/1475-7516/2004/03/004
[arXiv:hep-th/0312182 [hep-th]].

\bibitem{Jerome}
J.~Quintin, R.~H.~Brandenberger, M.~Gasperini and G.~Veneziano,
``Stringy black-hole gas in \ensuremath{\alpha}'-corrected dilaton gravity,''
Phys. Rev. D \textbf{98}, no.10, 103519 (2018)
doi:10.1103/PhysRevD.98.103519
[arXiv:1809.01658 [hep-th]].

\bibitem{Blau}
M. Blau, ``Lecture Notes on General Relativity''
(http://www.blau.itp.unibe.ch/GRLecturenotes.html).

\bibitem{Abramo}
V.~F.~Mukhanov, L.~R.~W.~Abramo and R.~H.~Brandenberger,
``On the Back reaction problem for gravitational perturbations,''
Phys. Rev. Lett. \textbf{78}, 1624-1627 (1997)
doi:10.1103/PhysRevLett.78.1624
[arXiv:gr-qc/9609026 [gr-qc]];\\
L.~R.~W.~Abramo, R.~H.~Brandenberger and V.~F.~Mukhanov,
``The Energy - momentum tensor for cosmological perturbations,''
Phys. Rev. D \textbf{56}, 3248-3257 (1997)
doi:10.1103/PhysRevD.56.3248
[arXiv:gr-qc/9704037 [gr-qc]].

\bibitem{BRreview}
R.~H.~Brandenberger,
``Back reaction of cosmological perturbations and the cosmological constant problem,''
[arXiv:hep-th/0210165 [hep-th]].

\bibitem{entropy}
G.~Geshnizjani and R.~Brandenberger,
``Back reaction of perturbations in two scalar field inflationary models,''
JCAP \textbf{04}, 006 (2005)
doi:10.1088/1475-7516/2005/04/006
[arXiv:hep-th/0310265 [hep-th]];\\
G.~Marozzi, G.~P.~Vacca and R.~H.~Brandenberger,
``Cosmological Backreaction for a Test Field Observer in a Chaotic Inflationary Model,''
JCAP \textbf{02}, 027 (2013)
doi:10.1088/1475-7516/2013/02/027
[arXiv:1212.6029 [hep-th]];\\
R.~Brandenberger, L.~L.~Graef, G.~Marozzi and G.~P.~Vacca,
``Backreaction of super-Hubble cosmological perturbations beyond perturbation theory,''
Phys. Rev. D \textbf{98}, no.10, 103523 (2018)
doi:10.1103/PhysRevD.98.103523
[arXiv:1807.07494 [hep-th]];\\
V.~Comeau and R.~Brandenberger,
``Back-Reaction of Long-Wavelength Cosmological Fluctuations as Measured by a Clock Field,''
[arXiv:2302.05873 [gr-qc]].

\bibitem{BounceRev}
R.~H.~Brandenberger,
``Alternatives to the inflationary paradigm of structure formation,''
Int. J. Mod. Phys. Conf. Ser. \textbf{01}, 67-79 (2011)
[arXiv:0902.4731 [hep-th]];\\
R.~Brandenberger and P.~Peter,
``Bouncing Cosmologies: Progress and Problems,''
Found. Phys. \textbf{47}, no.6, 797-850 (2017)
[arXiv:1603.05834 [hep-th]].

\bibitem{BV}
 R.~H.~Brandenberger and C.~Vafa,
 ``Superstrings In The Early Universe,'' 
 Nucl.\ Phys.\ B {\bf 316}, 391 (1989).
 
\bibitem{us}
S.~Brahma, R.~Brandenberger and S.~Laliberte,
``Emergent cosmology from matrix theory,''
JHEP \textbf{03}, 067 (2022)
[arXiv:2107.11512 [hep-th]];\\
S.~Brahma, R.~Brandenberger and S.~Laliberte,
``Emergent metric space-time from matrix theory,''
JHEP \textbf{09}, 031 (2022)
[arXiv:2206.12468 [hep-th]].

 \bibitem{Vafa}
P.~Agrawal, S.~Gukov, G.~Obied and C.~Vafa,
  ``Topological Gravity as the Early Phase of Our Universe,''
  arXiv:2009.10077 [hep-th].

\bibitem{Nayeri}
 A.~Nayeri, R.~H.~Brandenberger and C.~Vafa,
``Producing a scale-invariant spectrum of perturbations in a Hagedorn phase of string cosmology,''
Phys. Rev. Lett. \textbf{97}, 021302 (2006)
[arXiv:hep-th/0511140 [hep-th]];\\
R.~H.~Brandenberger, A.~Nayeri, S.~P.~Patil and C.~Vafa,
  ``Tensor Modes from a Primordial Hagedorn Phase of String Cosmology,''
  Phys.\ Rev.\ Lett.\  {\bf 98}, 231302 (2007)
  [hep-th/0604126].
  
\bibitem{ChibMukh}
V. Mukhanov and G. Chibisov,
 ``Quantum Fluctuation And Nonsingular Universe. (In Russian),''
 JETP Lett.\  {\bf 33}, 532 (1981) [Pisma Zh.\ Eksp.\ Teor.\ Fiz.\  {\bf 33}, 549 (1981)].
 
\bibitem{Starob}
A.~A.~Starobinsky,
``Spectrum of relict gravitational radiation and the early state of the universe,''
  JETP Lett.\  {\bf 30}, 682 (1979)
  [Pisma Zh.\ Eksp.\ Teor.\ Fiz.\  {\bf 30}, 719 (1979)].
  
\bibitem{Fabio}
F.~Finelli and R.~Brandenberger,
``On the generation of a scale invariant spectrum of adiabatic fluctuations in cosmological models with a contracting phase,''
Phys. Rev. D \textbf{65}, 103522 (2002)
[arXiv:hep-th/0112249 [hep-th]].

\bibitem{Peter}
Y.~F.~Cai, R.~Brandenberger and P.~Peter,
``Anisotropy in a Nonsingular Bounce,''
Class. Quant. Grav. \textbf{30}, 075019 (2013)
[arXiv:1301.4703 [gr-qc]].

\bibitem{Ekp}
 J.~Khoury, B.~A.~Ovrut, P.~J.~Steinhardt and N.~Turok,
 ``The Ekpyrotic universe: Colliding branes and the origin of the hot big
 bang,''
 Phys.\ Rev.\ D {\bf 64}, 123522 (2001) [hep-th/0103239];\\
J.~Khoury, B.~A.~Ovrut, N.~Seiberg, P.~J.~Steinhardt and N.~Turok,
  ``From big crunch to big bang,''
  Phys.\ Rev.\ D {\bf 65}, 086007 (2002)
  [hep-th/0108187].
  
\bibitem{Durrer}
R.~Durrer and F.~Vernizzi,
``Adiabatic perturbations in pre - big bang models: Matching conditions and scale invariance,''
Phys. Rev. D \textbf{66}, 083503 (2002)
doi:10.1103/PhysRevD.66.083503
[arXiv:hep-ph/0203275 [hep-ph]].

\bibitem{Ziwei}
R.~Brandenberger and Z.~Wang,
  ``Nonsingular Ekpyrotic Cosmology with a Nearly Scale-Invariant Spectrum of Cosmological Perturbations and Gravitational Waves,''
Phys.\ Rev.\ D {\bf 101}, no. 6, 063522 (2020)
  [arXiv:2001.00638 [hep-th]];\\
  R.~Brandenberger and Z.~Wang,
  ``Ekpyrotic cosmology with a zero-shear S-brane,''
  Phys.\ Rev.\ D {\bf 102}, no. 2, 023516 (2020)
  [arXiv:2004.06437 [hep-th]].
 
\bibitem{swamp}
  H.~Ooguri and C.~Vafa, 
``On the Geometry of the String Landscape and the Swampland,'' 
  Nucl.\ Phys.\ B {\bf 766}, 21 (2007);\\
   G.~Obied, H.~Ooguri, L.~Spodyneiko and C.~Vafa,
  ``De Sitter Space and the Swampland,''
  arXiv:1806.08362 [hep-th].

 
\bibitem{TCC1}
A.~Bedroya and C.~Vafa,
 ``Trans-Planckian Censorship and the Swampland,''
JHEP {\bf 2009}, 123 (2020)
  [arXiv:1909.11063 [hep-th]].
 
\bibitem{TCC2}
A.~Bedroya, R.~Brandenberger, M.~Loverde and C.~Vafa,
 ``Trans-Planckian Censorship and Inflationary Cosmology,''
 Phys.\ Rev.\ D {\bf 101}, no. 10, 103502 (2020)
 [arXiv:1909.11106 [hep-th]].

\bibitem{RHBrevs}
R.~Brandenberger,
  ``Fundamental Physics, the Swampland of Effective Field Theory and Early Universe Cosmology,''
  arXiv:1911.06058 [hep-th];\\
  R.~Brandenberger,
  ``Trans-Planckian Censorship Conjecture and Early Universe Cosmology,''
  arXiv:2102.09641 [hep-th];\\
R.~Brandenberger,
``String Cosmology and the Breakdown of Local Effective Field Theory,''
[arXiv:2112.04082 [hep-th]].  
 
\bibitem{Bernardo}
H.~Bernardo, S.~Brahma and M.~M.~Faruk,
``The inheritance of energy conditions: Revisiting no-go theorems in string compactifications,''
[arXiv:2208.09341 [hep-th]];\\
H.~Bernardo, S.~Brahma, K.~Dasgupta, M.~M.~Faruk and R.~Tatar,
``Four-Dimensional Null Energy Condition as a Swampland Conjecture,''
Phys. Rev. Lett. \textbf{127}, no.18, 181301 (2021)
doi:10.1103/PhysRevLett.127.181301
[arXiv:2107.06900 [hep-th]].

\end{thebibliography}
\end{document}